\documentclass[useAMS,usenatbib,usegraphics]{mn2e}
\usepackage{amssymb}
\usepackage{graphicx}
\title{Implementing TOPbase/Iron Project: Continuous Absorption
from Fe\,{\sc ii}}
\author[Charles R. Cowley and Manuel Bautista]
       {Charles R. Cowley$^{1}$\thanks{E-mail: cowley@umich.edu}
        and Manuel Bautista$^{2}$\thanks{E-mail:
        mbautist@ivic.ve}\\
$^{1}$Department of Astronomy, University of Michigan, Ann Arbor, MI
        48109-1090, USA\\
$^{2}$ IVIC, Laboratorio de Fisica Computacional, Apdo 21827,
        Caracas 1020A, Venezuela}

\date{Accepted}
\begin{document}
\pagerange{\pageref{firstpage}--\pageref{lastpage}} \pubyear{2002}

\maketitle

\label{firstpage}

\begin{abstract}

We discuss implementation of TOPbase and Iron Project opacities
for stellar spectral codes.  We use a technique employed by Peach,
where a Boltzmann-averaged cross section is calculated for
selected temperatures, and the opacity obtained from double
interpolation in temperature and wavelength.  It is straightforward
to include {\it all} levels for which cross sections have been calculated.
Boltzmann-averaged cross sections for Fe {\sc ii} show a local maximum
between 1700 and 2000\AA.  We suggest this feature arises from 
$3d^54snl \to 3d^54pnl$ transitions within Fe {\sc ii}.
IUE spectra of iron-rich CP stars show
local minima in this region.  Theoretical calculations of a
representative stellar continuum demonstrate that Fe {\sc ii}
photoionization contributes significantly to the observed
minima.

\end{abstract}


\section{Introduction and Rationale}

     State of the art Photoionization cross sections have
been available from The Opacity Project (Seaton 1994,
The Opacity Project Team, 1995) for 
more than a decade.  Nevertheless, stellar atmospheres codes 
have been slow to fully incorporate these results.   
Surely,  one reason for 
the delay in adopting these calculations is that 
metal opacities are not particularly important for many
stellar atmospheres.   The well-known opacities from hydrogen,
bound-free, and H$^-$, usually dominate other sources.

But it is also tedious to
incorporate the detailed TOPbase results into workable 
subroutines with typical algorithms, which treat levels
individually, or in groups.  This technique is difficult
for complex atoms, where no simple parameterization can
describe all of the cross sections adequately. 
In this paper, we propose adoption of the method
used by Peach (1967), which provides
a practical solution to this
problem.
We then discuss an example of stellar continuum
calculations that uses TOPbase/Iron Project
cross sections along with her algorithm.  We first 
outline the general problem of dealing with the
TOPbase/Iron Project material.

     Photoionization cross sections in TOPbase are given for 
typically hundreds of levels.   For each level, the cross 
sections may be given several hundred to a thousand photon 
energies.  The cross sections include large deviations from 
the smooth hydrogenic functions, that are possible in multi-electron 
systems.  
The resulting features may be quite sharp, being
essentially spectral lines, or they can be broader, when 
there is strong interaction between bound and continuum 
states.  Not only do Fano profiles appear, but also     
broad, deep absorption minima (Cooper minima), which    
result from nodes in the radial wave functions.  These are 
especially apparent in transitions involving $s$-electrons, 
as in Na I.  Broad absorption maxima can result from 
photoexcitation of core electrons (PEC), essentially 
bound-bound transitions of inner electrons.  These maxima
can appear in the cross sections of numerous levels.

     A consensus has emerged that it is permissable to smooth 
over the resonances of photoionization cross sections (cf. 
Bautista, Romano, and Pradhan 1998,  Prieto et al. 2002).  
This smoothing affects the 
sharp features which we shall call {\it high-frequency 
components}, but leaves broader ones, such as Cooper minima 
or PEC.  Prieto has published a number of
resonance-smoothed photoionization cross sections (RSP) on 
his website:
\newline http://hebe.as.utexas.edu/at/at.cgi
Our current opacity routines use the Prieto
RSP's for all atoms and ions with the exception of Fe {\sc i} 
and {\sc ii}.
Specifically, the following atoms and atomic ions are included:
Li {\sc i}, Be {\sc i} and {\sc ii}, B {\sc i} and {\sc ii}, 
C {\sc i},
Na {\sc i}, Mg {\sc i} and {\sc ii}, Al {\sc i}, 
Si {\sc i} and {\sc ii}, S {\sc i}, and 
Ca {\sc i} and {\sc ii}.

For Fe {\sc i}, we have
sampled each 100\AA \ from material discussed by 
Bautista (1997).  The
Fe {\sc ii} cross sections are from a web site of Dr.
Nahar
\newline http://www-astronomy.mps.ohio-state.edu/~nahar/,
cf. Nahar and Pradhan 1994).

While the RSP's omit some of the higher TOPbase
levels, they may still contain several hundred levels (e.g. 
Si {\sc i}), with several hundred data points for each level.  If 
all of this information were included in a standard FORTRAN 
data statement, the result would be thousands of records 
long.  However, for practical calculations, the individual
cross sections are not necessary.  We need only their appropriately
weighted sums.

\section{Algorithm and calculations}
  
    The continuous opacity due to a given atom or atomic ion 
is the weighted sum of the cross sections over a number of levels.  We 
define the K-factor:

\begin{equation}
Kfact = \sum_n g_n\cdot 
\exp{[-(\chi_n/kT)]}\cdot\sigma_n(\lambda)
\end{equation}

\noindent Here, $g_n$ is the statistical weight of the 
level $n$, $\chi_n/kT$ the appropriate Boltzmann factor, 
and $\sigma_n(\lambda)$ is the photoionization cross 
section. 

Consider the opacity due to photoionizations from level $n$
of
an atom or atomic ion in the $i^{th}$ ionization stage.  Write
the corresponding number density as $N_{n,i}$.  Put
$N_{t,i}$ for the sum of the $N_{n,i}$'s in the $i^{\rm
th}$ stage of ionization.  We have 

\begin{equation}
N_{n,i} = N_{t,i}{g_n\over u_i(T)}\cdot \exp(-\chi_n/kT),
\end{equation}

\noindent where $u_i(T)$ is the partition function for the 
$i^{\rm th}$ ionization stage.  The opacity due to all 
levels {\it from this ion} is thus

\begin{equation}
\varkappa_i(\lambda) = N_{i,t}{1\over u_i(T)}\cdot Kfact.
\end{equation}

In LTE, this K-factor is a function only of the temperature 
and wavelength.  Since it is a sum over cross sections, 
some of the high-frequency structure for individual levels
will be smoothed.  One sees primarily, the absorption edges,
and the most pronounced structure, for example, the PEC.
The algorithm consists of computing this K-factor over the
wavelengths likely to be or relevance, say from 500 to 
8000A, for a relatively small number of temperatures.  
It is straightforward to include {\it all} of the levels in
TOPbase, or the RSP's.  

The opacities for any given atom or atomic ion are then 
found from bilinear interpolations in this 
temperature-wavelength grid.
                            
We have calculated the K-factor at 20\AA \ intervals,
for all species except for Fe {\sc i}.

With 20\AA \ sampling, K-factors were first obtained for 376 points
[(8000-500)/20 + 1].  Then, unnecessary points were eliminated
wherever ``interior'' data could be fit by linear
interpolations from bounding points to within 2\%.  For
example, if the data at $\lambda$4020 could be fit by
interpolation between the points at $\lambda\lambda$4000
and 4040, the point at $\lambda$4020 was dropped.  This
technique resulted in considerable shortening of the
data statements, especially for lower tempertures.

The current routines for
Fe {\sc i} and {\sc ii} are considered provisional, because they have
not been resonance smoothed.
The Nahar Fe {\sc ii} photoionization file is some 33
megabytes in size, and contains cross sections for 745 
states (or ``$LS$-multiplets'').  We computed K-factor's 
every 20\AA\,  ($\lambda\lambda$500--8000) for 8 temperatures
(2000 - 16000K).  The adopted cross section at each of the
grid wavelengths is an average of the nearest four TOPbase
points.  We have postponed improving this algorithm until
smoothed cross sections are available.  The current
Fe {\sc ii} subroutine
is some 360 FORTRAN statements, including data statements,
comments, and commented diagnostics.
Comparisons between
the K-factors for smoothed and unsmoothed cross sections
for lighter species show differences in detail, but not
major trends, extending over tens of angstroms.  The
feature discussed below is relatively broad.

\begin{figure}
\includegraphics[width=84mm]{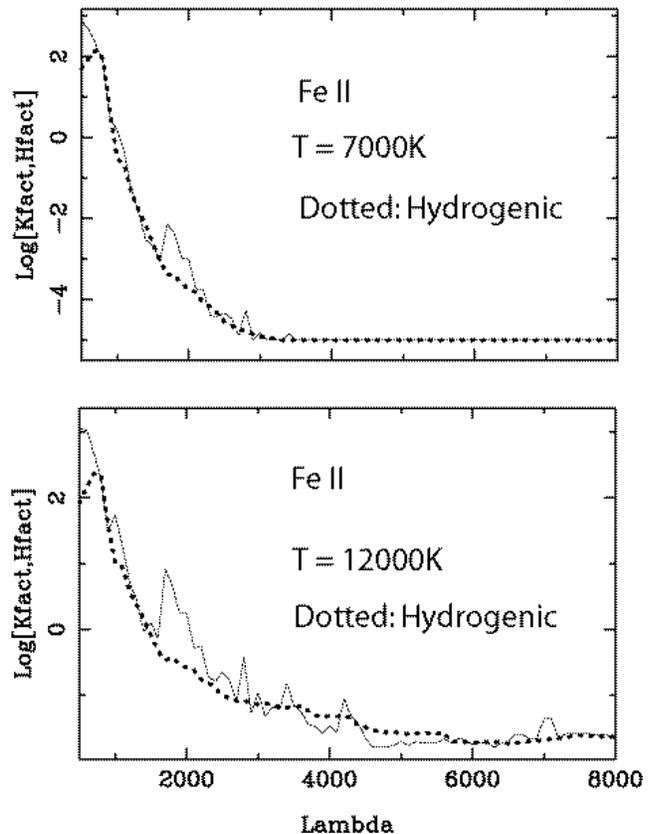}
\caption{K-factors for Fe {\sc ii}}
\label{kfact4.lab}
\end{figure}

\begin{figure*}
\includegraphics[width=176mm]{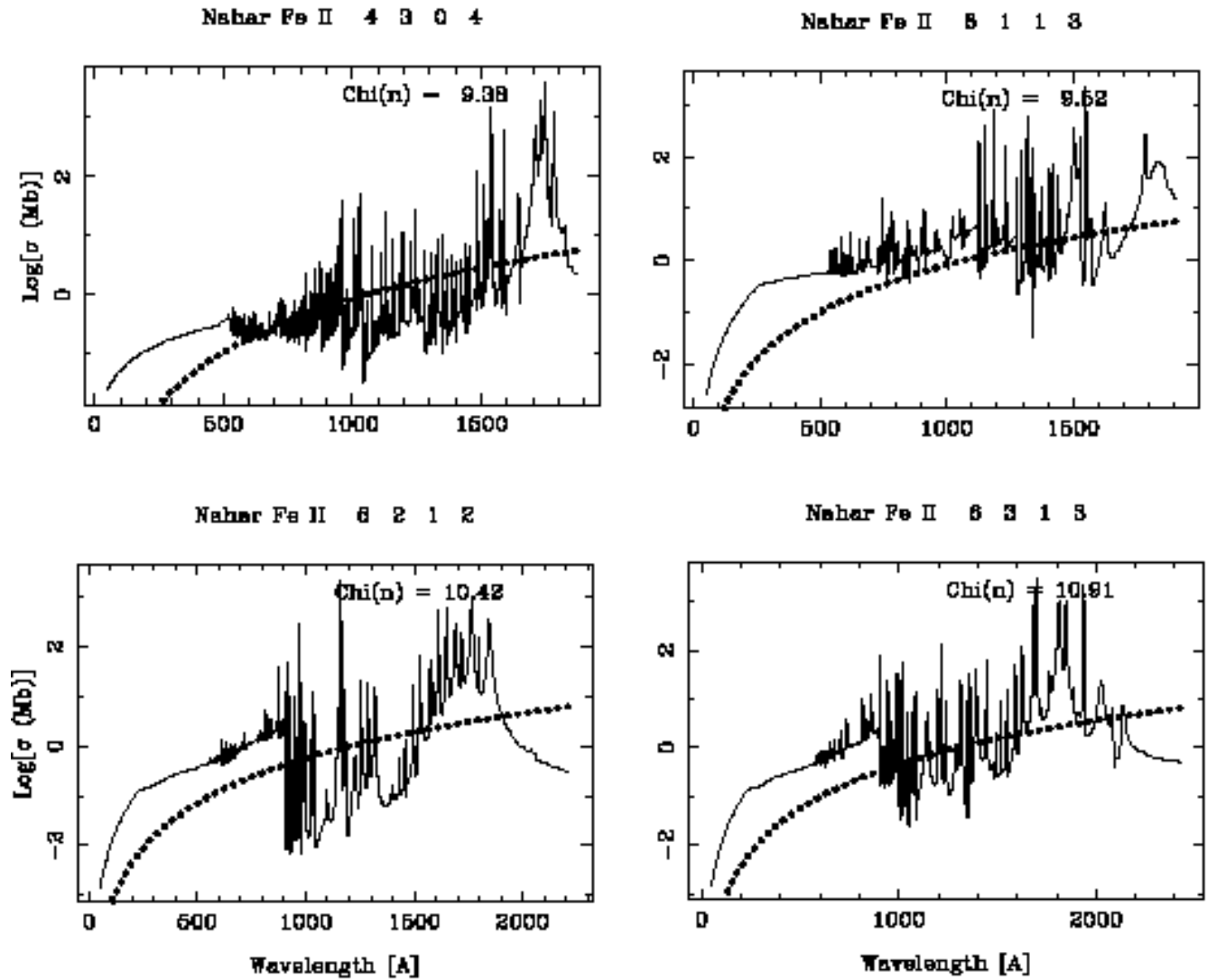}
\caption{Fe {\sc ii} cross sections (Nahar and Pradhan 1996) for
levels with wide maxima in the region 1700--2000\AA.  Hydrogenic
cross sections are shown by the dotted curves.
The
first three digits at the top of each plot 
identify the level or term by multiplicity,
$L$-value, parity (0=even,1=odd).  The fourth digit is a
sequential number, from low to high excitation, of the term
described by the first three digits.  For example, 4 3 0 4
describes an even $\rm ^4F$ term, that is the fourth such
term in order of excitation.  In TOPbase, the first three
digits are called ISLP and the fourth, ILV. Chi(n) is he excitation
energy of the level in eV.}
\label{xsec4.lab}
\end{figure*}

\begin{figure}
\includegraphics[width=84mm]{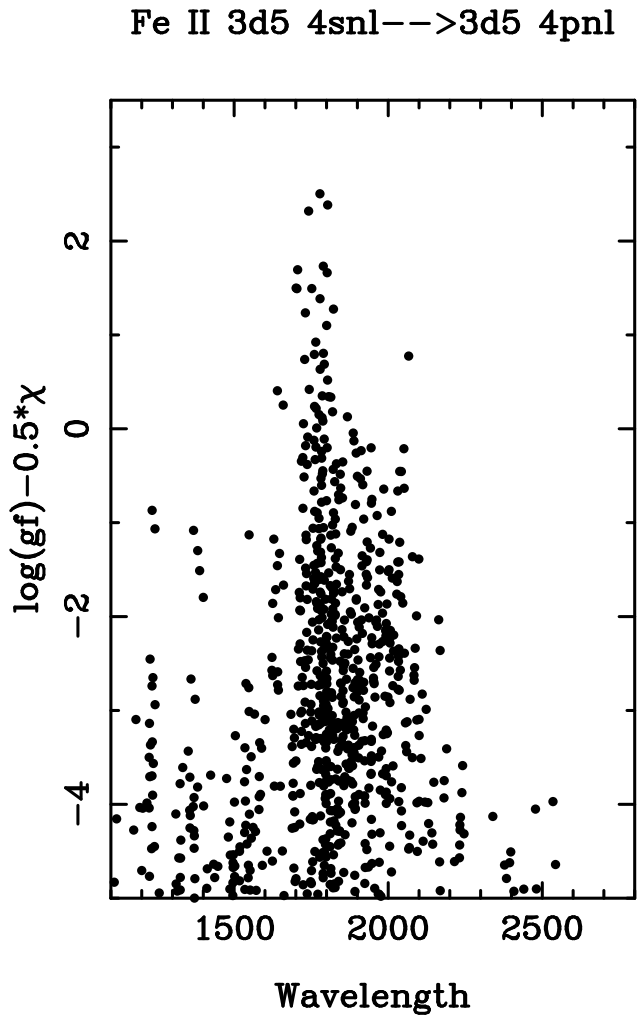}
\caption{Intensity factors for Fe\,{\sc ii} lines subject to autoionization.
The ordinate, $\log(gf)-0.5\cdot\chi$, is roughly proportional to 
the logarithm of the strength of an absorption line if saturation did 
not occur.  The logarithmic Boltzmann factor, corresponds to 
$\theta = 5040/T = 0.5$ or 1080K.}   
\label{res.lab}
\end{figure}

Fig.~\ref{kfact4.lab} shows plots of the K-factors for 
Fe {\sc ii} at 7000
and 12000K, based on the Nahar data file.  The dotted line
shows the resulting ``H-factors,'' 
that is, where the photoionization cross
sections are all assumed hydrogenic (Cowan 1981, Eq. 18.47). 
The feature between about 1700 and 2000\AA\, is notable.  We
suggest that it 
arises from core photoexcitations $3d^54snl \to 3d^54pnl$ within Fe {\sc ii}. 
It is important to notice that in these sort of transitions the outer
$nl$ electron acts as spectator and has only little effect on the wavelengths
and radiative rates of the core transitions.
The related $3d^54s \to 3d^54p$
transitions in Fe {\sc iii}
give rise to the strong lines in UV Multiplets 34, 50, 51,
and 52.  
We have made a further investigation of the nature of the $3d^54snl \to
3d^54pnl$
transitions in Fe\,{\sc ii} using the atomic structure code
{\sc autostructure}
(Badnell 1986, 1997; Badnell \& Pindzola
1989).   This code is an extension of the atomic structure program
{\sc superstructure} by (Eissner et al. 1974). It allows for
the calculation
of level energies, and  radiative and Auger rates within
a statistical Thomas--Fermi--Dirac model
potential (Eissner \& Nussbaumer 1969).
Thus, we compute the total of 8988 $LS$-coupling dipole allowed
transitions
$3d^54snl \to 3d^54pnl$ for $5\le n \le 10$ and $0\le l \le 3$, with
the majority of the $3d^54pnl$ states lying above the the first
ionization threshold. A little
less than half of these transitions, 4090, end up in upper states
that can autoionize
to Fe\,{\sc iii} target states.  Such transitions manifest
themselves
as PEC resonances in the close coupling photoionization cross sections
of
Nahar and Pradhan (1994). 
Several of the cross sections that contribute to
the local peak in the 1700 to 2000\AA\, region are shown
in Fig.~\ref{xsec4.lab} (following page).
The remaining 4898 transitions go to levels
that
do not autoionize and must be treated as bound-bound transitions.
We show a subset of the former lines in Fig~\ref{res.lab}.
The
concentration of transitions in the region 1700-2000\AA\, is clear.

\section{Observations}

\begin{figure*}
\includegraphics[width=176mm]{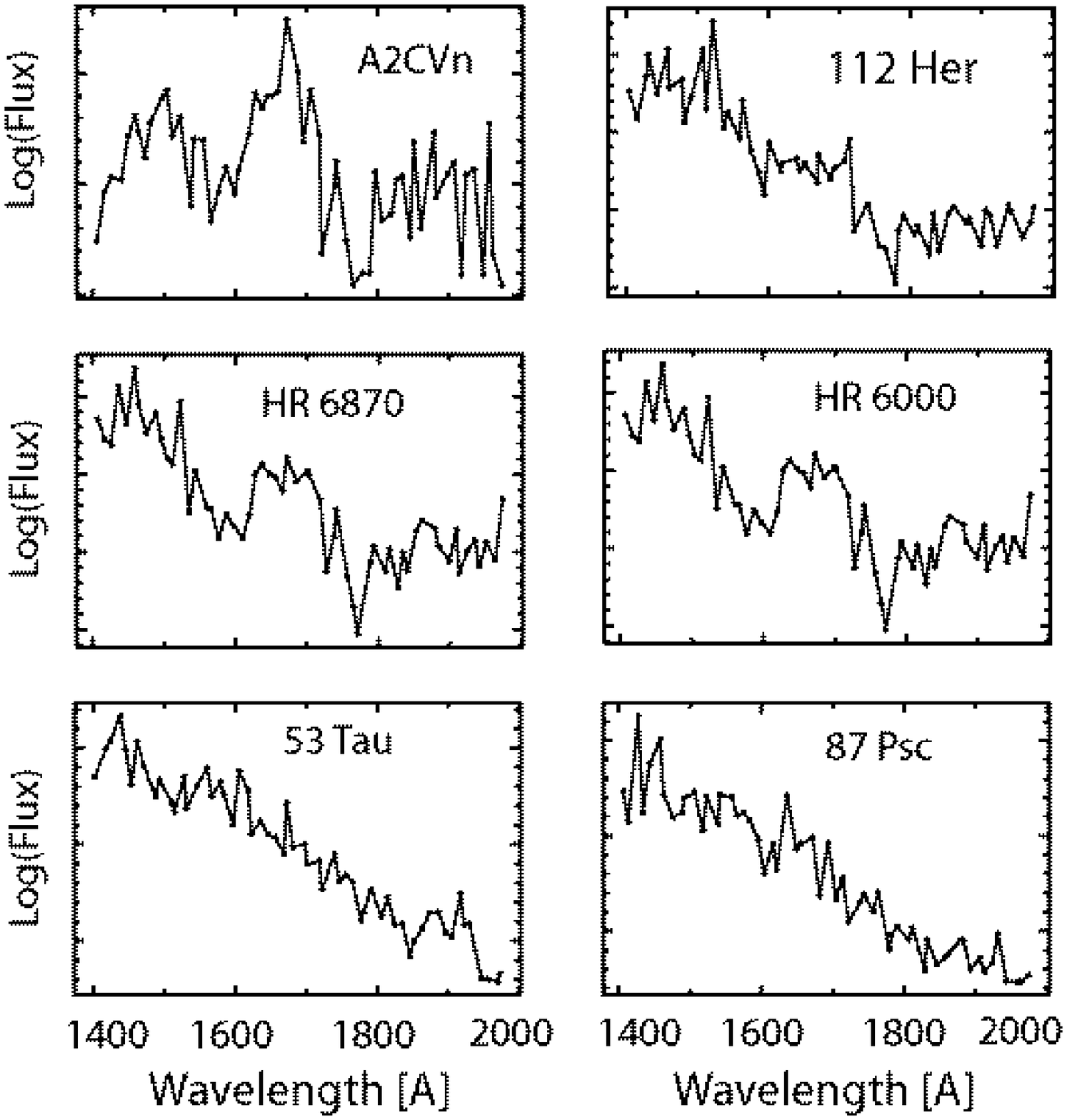}
\caption{The top four panels are IUE spectra of iron-rich
stars.  The lower two panels show stars with iron deficiencies.
Only the highest points within 10\AA\, intervals are plotted. The
original spectra are available on line at the site:
http://archive.stsci.edu/iue/.}
\label{star4.lab}
\end{figure*}


IUE spectra of iron-rich stars show significant depressions
in the wavelength range from roughly 1700-1800\AA, as is
shown in the top 4 panels of Fig.~\ref{star4.lab}.
These spectra were
downloaded from the url
http://archive.stsci.edu/iue/.  Because of the high noise
level of the spectra, we have chosen to plot only the maxima 
within 10\AA \ regions.  Complete spectra may be seen at the
url cited in the figure.  
Smith and Dworetsky (1993)
report high iron abundances in 112 Her, and HR 6000.
The magnetic Ap star $\alpha^2$CVn has long been known to be
iron rich (cf. Cohen 1970), while Muthsam and Cowley (1984)
showed HR 6870 (HD 168733) to be quite iron rich. The stars
87 Psc and 53 Tau are iron poor (see Smith and
Dworetsky 1993),  and their spectra lack the strong minima
seen in the other four stars.

Lanz, et al. (1996) discuss the region
on the short wavelength side of 1600\AA.  The interpretation is
not straightforward, though the figures shown here are clearly
relevant for any interpretation.
Note that HR 6000, notorious for its lack of strong
Si\,{\sc ii} lines in ground-based photographic spectra, also lacks
the depression short of $\lambda$1600 (cf. Andersen, et al.
1984).

\section{Application to Fe \,{\sc ii}}

\begin{figure}
\includegraphics[width=74mm]{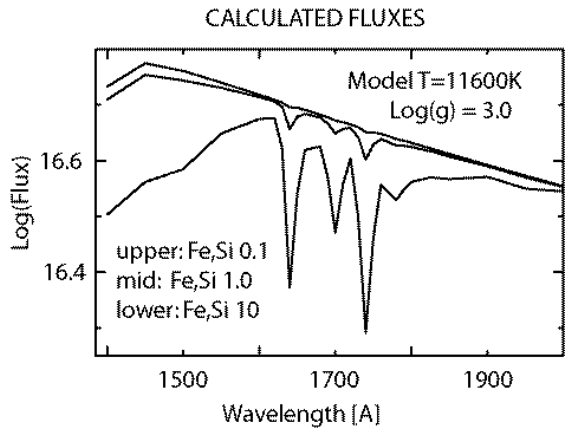}
\caption{Calculated continuum fluxes for $T_e = 11600$K,
and $\log(g) = 3$.  The plots, from upper to lower show
results for iron and silicon 0.1--, 1.0--, and 10.0--times
solar abundances}
\label{compfl4.lab}
\end{figure}

Fig.~\ref{compfl4.lab} shows calculated spectra in the wavelength region
under consideration.  The model was based on the adopted
parameters of Kochukhov et al. (2002), an effective temperature
of 11600K and log(g) = 3.0.  We also followed these authors
in using crude abundances to set the model.  We took abundances
0.1 SAD (Grevesse and Sauval 1998) for elements helium through
neon, and 10 SAD for all heavier elements.  This was only for
the determination of the temperature-pressure structure.  A
flux-constant model was derived from an Atlas 9 model (Kurucz 1993),
but the depth depencence was recalculated using Michigan routines
as described in earlier papers (cf. Cowley 1996), except that our opacity
routines now incorporate TOPbase data for the species enumerated
above.

The three curves were all based on the same $T(\tau)$, though a
new pressure structure was calculated based on the different iron
and silicon abundances.  Thus, the flux is not strictly constant
in the models without the enhanced abundances.  We do not consider
this significant in the present exploratory calculations.  The
three curves shown are for iron and silicon enhanced by a factor
of 10 (lower), solar (middle), and depleted by a factor of 10 (upper).
All elements other than iron and silicon were assumed to have the
same abundances used in the Atlas9 model.

The calculations show clearly that the Fe\,{\sc ii} continuum is significant
in this region.  Of the three calculated minima, the one at the
shortest wavelength (about $\lambda$1640) is not well matched to
the observations.  It is certainly Fe\,{\sc ii} and not Si, as a separate
calculation (not shown) has revealed.  There is a possibly corresponding
feature to be seen in the spectra of $\alpha^2$CVn and HR 6000,
though much weaker than in the calculations.  This region deserves
further study.

The two longer-wavelength minima, at
approximately $\lambda\lambda$1700 and 1740 fit the observations
of the iron-rich spectra reasonably well.  There is a small discrepancy
in the wavelengths; the calculated minima are some 20 to 30\AA \ short
of the stellar minima.  But these discrepancies are 2\% or less,
well within the accuracy with which energy levels in Fe\,{\sc ii}  were
calculated (cf. Nahar and Pradhan, op. cit. Table 1).

\section{Conclusions}

Some of the absorption in the region $\lambda\lambda$1700--2000 of the
spectra shown in Fig.~\ref{star4.lab} will result from the strong
transitions in Fe\,{\sc ii} as well as 
the multiplets mentioned above in Fe\,{\sc iii}.  But we show
here that a measurable contribution to the depression must be due
to the specific shapes of the Fe\,{\sc ii} photoionization cross sections.
Note that the lowest of the levels shown in Fig.~\ref{xsec4.lab}, at
9.38eV is the 76th level above ground.  Any attempt to treat higher
levels in Fe\,{\sc ii} by a hydrogenic approximation or simple
parameterization would miss the relevant structure.     

A number of loose ends need to be tidied before our metal opacity
routines can be considered satisfactory.  We need smoothed cross sections
for Fe\,{\sc i} and \,{\sc ii}, as well as other heavy species, 
such as Mn\,{\sc i} and \,{\sc ii}
or Co\,{\sc i} and \,{\sc ii} that can be relevant for special stars.  
It is also
necessary to make sure that the partition functions used with the
TOPbase/Iron Project cross sections are compatible with those levels.
Provisional, spot checks for a few spectra show that this is not 
likely to be a serious problem, as the calculated levels are usually
within a few per cent of the laboratory values.  Line opacity was not
relevant for the current calculations.  Eventually, we must consider 
the overlap between the TOPbase ``resonances'' and
predicted lines in the Kurucz data base.

These are matters for future contributions. 

\section*{Acknowledgments}

We thank D. J. Bord,  F. Castelli, R. D. Cowan,
R. L. Kurucz, S. Johansson,
L. L. Lohr, S. Nahar, A. K. Pradhan, and 
M. J. Seaton for various comments and
advice.  We join many astronomers
grateful to the workers at the IUE Project and Space Telescope
Multimission-Archive for providing observational material in
such convenient and useful forms.     
Apologies are surely due to others whose names were
inadvertently omitted.

\label{lastpage}

\end{document}